# Multidimensional sound propagation in 3D high-order topological sonic insulator

Fei Meng[1,2], Yafeng Chen[1,3], Weibai Li[1], Baohua Jia[1,*], Xiaodong Huang[1,*]

1 Centre of Translational Atomaterials (CTAM), Faculty of Science, Engineering and Technology, Swinburne University of Technology, Hawthorn, VIC 3122, Australia

2 State Key Laboratory of Geomechanics and Geotechnical Engineering, Institute of Rock and Soil Mechanics, Chinese Academy of Sciences, Wuhan, 430071, China

3 Key Laboratory of Advanced Technology for Vehicle Body Design & Manufacture, Hunan University, Changsha 410082, China

* Corresponding author: Baohua Jia, bjia@swin.edu.au; Xiaodong Huang, xhuang@swin.edu.au

**Abstract:** High-order topological insulators (TIs) develop the conventional bulk-boundary correspondence theory and rise the interest in searching innovative topological materials. To realize a high-order TI with a wide passband of 1D and 2D transportation modes, we design non-trivial and trivial 3D sonic crystals whose combination mimics the Su-Schrieffer-Heeger model. The high-order topological boundary states can be found at the interfaces, including 0D corner state, 1D hinge state, and 2D surface state. The fabricated sample with the bent two-dimensional and one-dimensional acoustic channels exhibits the multidimensional sound propagation in space, and also verifies the transition between the complete band gap, hinge states, and surface states within the bulk band gap. Among them, the bandwidth of the single-mode hinge state achieves a large relative bandwidth 9.1%, in which sound transports one-dimensionally without significant leak into the surfaces or the bulk. The high-order topological states in the study pave the way for multidimensional sound manipulation in space.

Condensed matter physics has undergone a deep revolution after the propose of topological phases of matter [1, 2]. Topological insulators (TIs) are materials that behave as insulators in the bulk but conductors on its edges. Normally, two-dimensional (2D) TIs host topologically protected one-dimensional (1D) edge states, while three-dimensional (3D) TIs have topologically protected 2D surface states. Recently, a new breed of topological phase — high-order topological insulators has been suggested [3-10]. A higher-order TIs in $d$ dimensions possess ($d$-1)D or ($d$-2)D edge states, which obey the conventional bulk-boundary correspondence.

High-order TIs have been investigated in mechanical metamaterials [11-13], electrical circuits

[14, 15], optical waveguides [9, 16], photonic crystals [17-19], and sonic crystals [20-23]. Among these, sonic crystals have received much attention. Sonic crystals are structural materials constructed by periodically arranged solid or fluid scatters/resonators. Since their structures in macroscopic can be designed and fabricated as almost any shape, the desired energy bands can be formed conveniently. Unlike electronic systems, the absence of Fermi level makes the entire spectrum easily accessible. Therefore, sonic crystals have become an ideal platform for studying the topological phases of matter [20-32].

Currently, the researches on high-order TIs mainly focused on 2D materials/metamaterials, which are periodic in two dimensions and homogeneous in the third. Comparing to their 2D counterparts, 3D TIs provide one more dimension in space to harness the topological phases of matters, and use them to manipulate the propagation of current, light, or sound. Theoretically, 3D High-order TIs enable to possess topologically protected 2D surface states, 1D hinge states, and zero-dimensional (0D) corner states [4, 6, 8]. The corner states enable applications such as cavity [19], while hinge and surface states make the transportation of matter/signal in various dimensions possible. Recently, researchers have observed 1D hinge states in bismuth [33], and 0D corner state in sonic crystals [20, 34]. However, the high-order topological states of 3D sonic crystals often bury in the surface or bulk states and the low dimensional transportation modes are thus hard to access separately [34]. The realization of a 3D sonic insulator with a wide bandwidth of independent transportation modes is still unreported, although the separation of hinge and surface states is meaningful in practical applications.

In this paper, we elaborately design a third-order TI based on 3D trivial and nontrivial sonic crystals, which can be viewed as a physical realization of the Su-Schrieffer-Heeger (SSH) model. Those two specially designed sonic crystals have a wide overlapped bulk band gap so that the constructed TI can readily hold the 0D corner states, 1D hinge states, and 2D surface states simultaneously. Both the simulation and experiment successfully demonstrate that sound propagates along spatial polygonal lines under 1D hinge states and bent interfaces under 2D surface states. In the single-mode hinge state passband, sound propagates like in an "acoustic fiber". The hierarchy and transition of multidimensional sound propagation are directly detected in experiments.

To design a high-order sonic TI, it needs to create sonic crystals with trivial and non-trivial topological properties, respectively. Meanwhile, a wide overlapped bulk band gap is necessary so that TI can host the topological propagation modes over a wide frequency range. Inspired by our previous work on topology optimization of 3D phononic and sonic crystals [35, 36], two novel structures of sonic crystals are designed out as illustrated in Figure 1a and 1b. The non-trivial sonic crystal (Figure 1a) is constructed by removing a pyramid from each face of a solid cubic (as indicated in the insets). Consequently, six air cavities are formed and coupled by the ports between them. The trivial sonic crystal (Figure 1b) is constructed similarly but subtracts one more pyramid on each face. Then, the center of unit cell shifts by ($0.5a$, $0.5a$, $0.5a$) to induce the switch of eigenmodes' parity on high-symmetry point X. The unit cells have a simple cubic lattice with a side length $a$ ($a$ = 16mm). The energy bands of these two sonic crystals are given in Figure 1c and 1d. The two band gaps have a relative size of 42.3% and 65.8%, respectively, and overlap between 10.34 kHz to 15.88 kHz.

The above nontrivial and trivial sonic crystals follow the rules of SSH model in 3D [37-39]. Berry curvature vanishes everywhere in the first Brillouin zone (BZ) [40, 41] due to the geometric symmetries of simple cubic lattice and the time-reversal symmetry. The topological properties of these sonic crystals can be characterized by the vectorial Zak phase or fractional bulk polarization. The 3D bulk polarization is the integration of the Berry connection over the momentum space [40, 42].

$$\mathbf{P} = -\frac{1}{(2\pi)^3} \iiint dk_x dk_y dk_z \mathrm{Tr}[\mathbf{A}_n(\mathbf{k})] \tag{1}$$

$$\mathbf{A}_n(\mathbf{k}) = i\langle u_n(\mathbf{k})|\partial_\mathbf{k}|u_n(\mathbf{k})\rangle$$

where $\mathbf{k} = (k_x, k_y, k_z)$ is the wavevector. $n$ refers to the band index which runs over all the bands below the band gap. $\partial_\mathbf{k}$ is the vector gradient operator in $\mathbf{k}$-space. $|u_n(\mathbf{k})\rangle$ is the periodic part of the Bloch wave function. $\mathbf{P} = (P_x, P_y, P_z)$ is the 3D bulk polarization and the location of Wannier center. Because of the full octahedral symmetry $O_h$ of the simple cubic lattice, we have $P_x = P_y = P_z$. The integration in Eq.1 is conducted over the first Brillouin zone (BZ). The inversion symmetry of the simple cubic lattice quantizes $P_i$ to either 0 or 1/2. The quantization value can be determined by the parities of Bloch eigenstates at the high-symmetry points in the BZ [40, 42]:

$$P_i = \frac{1}{2}(\textstyle\sum_n q_i^n \mod 2) \tag{2}$$

$$(-1)^{q_i^n} = \frac{\eta_n(X_i)}{\eta_n(\Gamma)}$$

The summation is taken over all the bands below the band gap. $i = x, y, z$ represents the direction. $X_i$ is the corresponding high symmetric point X ($\mathbf{k} = (a/\pi, 0, 0)$), Y ($\mathbf{k} = (0, a/\pi, 0)$), and Z ($\mathbf{k} = (0, 0, a/\pi)$). $\eta_n$ is the parity of the $n^{th}$ band at this point. The parities of eigenstates at $\Gamma$ ($\mathbf{k} = (0, 0, 0)$), and X are denoted in Figure 1c and 1d. "+" indicates an even parity while "−" indicates an odd parity. Note that the parities of Y, Z are the same with X due to the mirror symmetries of simple cubic lattice, form Equation 2 we can get $\mathbf{P} = (0, 0, 0)$ for the trivial sonic crystal, which denotes a trivial topological phase; $\mathbf{P} = (1/2, 1/2, 1/2)$ for the non-trivial sonic crystal, indicating a non-trivial topological phase. The Wannier center locates at the corner of the unit cell implies a third-order topology, and topological surface, hinge and corner states arises [34].

Multidimensional topological states including high-order corner and hinge states can be realized on the specific boundaries between trivial and non-trivial sonic crystals. In the following, we numerically reveal the existence of the 1D and 2D propagation states, while the analysis of 0D corner states is shown in the *Supporting Information, Section 1*. To investigate the 2D surface state, a numerical model of a ribbon-like supercell consists of 8 trivial unit cells and 8 non-trivial unit cells is built up as shown in Figure 2a. The projected band diagram in $k_x$-$k_y$ plane of the supercell is illustrated in Figure 2b. The projected bulk band gap is between 10.38~15.90 kHz, and three in-gap bands arise. The sound pressure fields of the in-gap states at M are depicted in Figure 2c. It can be seen that the high-pressure region localizes at the interface, which indicates the 3 in-gap bands are surface states. The antisymmetric dipolar pressure fields occur at the two lower surface bands while the symmetric monopolar one at the highest surface band. Different from the gapless surface states in Chern insulator [43] or Spin-Chern insulator [44, 45], these surface bands are gapped, which is a feature of high-order topological insulators.

To analyze the 1D hinge state, a slab-like supercell is built as shown in Figure 3a. The inner part of the supercell is the trivial sonic crystal with 8×8 unit cells, which is surrounded by 4 layers of unit cells of the non-trivial sonic crystal. Four interfaces are formed and meet at the four hinges. The projected band diagram presented in Figure 3b shows the states on the dimensional hierarchy of bulk, surface, and hinge. The surface state emerges in the bulk band gap, and the four degenerated hinge state occurs in the surface band gap. The four hinge states at $k_z$ = 0, 14.02 kHz are illustrated in Figure 3c, in which the high-pressure region concentrates on one of the four hinges. The projected band diagram indicates the transition of propagation modes for different sound frequencies. In the bulk band gap, the surface states occupy 10.72~12.81 kHz and 14.43~15.89 kHz. From 11.96 to 12.81 kHz, surface states and hinge states coexist. A frequency range with pure hinge states is realized from 12.81 to 14.02 kHz, with a relative bandwidth (bandwidth to mid-band frequency ratio) of 9.1%. Sound waves in this single-mode hinge state region can propagate in the high-order TI via the hinges only. It can be used to achieve 1D sound transportation in 3D space, acting as an acoustic fiber.

Between the hinge states and upper surface states, a small band gap (14.02~14.43 kHz) for all propagation modes emerges. It is referred to as the complete band gap in this paper. The 0D corner states are found within it (see the *Supporting Information, Section 1*).

The experiment is conducted to confirm the existence of multidimensional topological sound propagation, especially the 1D hinge propagation. The fabricated sample is shown in Figure 4a. The trivial sonic crystal forms a 90° bent square tunnel, with a side length of 5*a*. The trivial sonic crystal is surrounded by 4 layers of non-trivial sonic crystals in three sides, while the bottom is sealed by a 5 mm thick slab. Hence two bent hinges and three surfaces are formed. The experimental setup is depicted in Figure 4b. Different from the approach in the previous study [20-23, 34] that put both excitation and probes inside the sample, we locate the sound source on one side and straightforwardly measure the sound signal that propagates through the sample. A speaker exerts the sound excitations on the left side of the sample through a square tube (side length 5*a*). On the right side of the sample, microphone probes are set at the exit of hinges and interfaces to measure the sound pressure. The bulk states are detected via a microphone at the center of the upper side of the sample.

The measured transmission spectra for sound ranging from 10 to 17 kHz are illustrated in Figure 4c. In the bulk band gap, the bulk probe detects almost no signal. In the frequency range of surface states (10.72~12.81 kHz and 14.43~15.89 kHz), both hinge probes and surface probes detect sound transports from the source. The spectra indicate that within the latter surface state range corresponding to the third surface band in Figure 2b, the sound has better transmission. The relative bandwidth of this surface state is about 9.6%. In range 12.81~14.02 kHz, i.e. the single-mode hinge state region, the hinge probe has a much higher response than those of the surface probes, indicating that the 1D propagation modes are dominant. The response of hinge probes decreases rapidly when the frequency approaches the complete band gap (14.02~14.43 kHz). In this complete band gap, almost no signal is detected by any probe. The transmission spectra agree well with our theoretical analysis of the topological boundary states in the previous section.

To verify the detected response indeed concentrate on the hinges and interfaces, we arbitrarily choose two frequencies, 13.30 kHz and 15.40 kHz, and measure the sound pressure field near the measure plane. The sound pressure distribution inside the sample is revealed by simulation as shown in Figure 5a, where the non-trivial sonic crystals are peeled off to give a close look of the topological interfaces. The position of the sound source is adjusted for 13.30 kHz to excite the two independent hinges separately. The simulated sound pressure fields at 1 mm away from the measure plane are shown in Figure 5b. In the experiment, the sound pressure fields at the same position are scanned and plotted in Figure 5c, which agree well with the simulation results. At 13.30 kHz, the peaks of sound pressure concentrate at the exits of hinge 1 and hinge 2, and decay rapidly into the bulk and surface. At 15.40 kHz, the acoustic energy is highly localized at the interfaces of the two sonic crystals, which clearly visualizes the dimensional hierarchy of topological propagation modes. For simplicity, the experiment only demonstrates the tunnel with one bend in *x-y* plane, but obviously, both the surfaces and hinges can be bent to any directions like a 3D strong TI (a more complex numerical model can be found in the *Supporting Information, Section 2*).

In conclusion, we create 3D topological trivial and non-trivial airborne sonic crystals with novel structures and wide bulk band gaps. The combination of them realizes a 3D acoustic analog of the SSH model and achieves the high-order topological phases at their interfaces. Numerical simulation reveals the 1D and 2D propagation modes, including a single-mode hinge state passband with a relative bandwidth of 9.1%. A sample with bent hinges and interfaces is fabricated and tested. The experiment demonstrates the multidimensional sound propagation and verifies the modes transition between the complete band gap, hinge states, and surface states. In the frequency range of the single-mode hinge state, sound transports one-dimensionally through a polyline path without significant leak into the surfaces or the bulk. The High-order TI realized in this research has applications in local acoustic field enhancement, multidimensional acoustic manipulation, and acoustic sensing or probing.

**Experimental Section**

*Numerical simulation*: The numerical simulations in this work are conducted by COMSOL Multiphysics. The density of air is taken as 1.21 kg·m$^{-3}$, sound speed in air is 343 m·s$^{-1}$. The interface between structure and air are treated as sound hard boundaries.

*Experiment*: The sample is made of photosensitive resin via 3D stereolithography (SLA) printing. The fabrication error is less than 0.1 mm. The resin's modulus is 2.8 MPa and density is 1.3 g·cm$^{-3}$. The sound source is a HIVI RT1C-A speaker. The sine wave sound signal is generated by the built-in sound card of BSWA MC3242 data collector. Sound pressure is picked up by NI 9233 data acquisition card with MPA416 microphones. For the field scanning experiment, the position of microphone is controlled by a motorized linear stage, and the space between two adjacent test points is 2mm,.

**Acknowledgment**

F. M. would like to acknowledge the support from the Open Research Fund of State Key Laboratory of Geomechanics and Geotechnical Engineering, Institute of Rock and Soil Mechanics, Chinese Academy of Sciences, Grant No. Z018009. X. H. wishes to acknowledge the financial support from the Australian Research Council (FT130101094).

**Keywords**

sonic crystals, high-order topological insulators, Su-Schrieffer-Heeger model, acoustic fiber

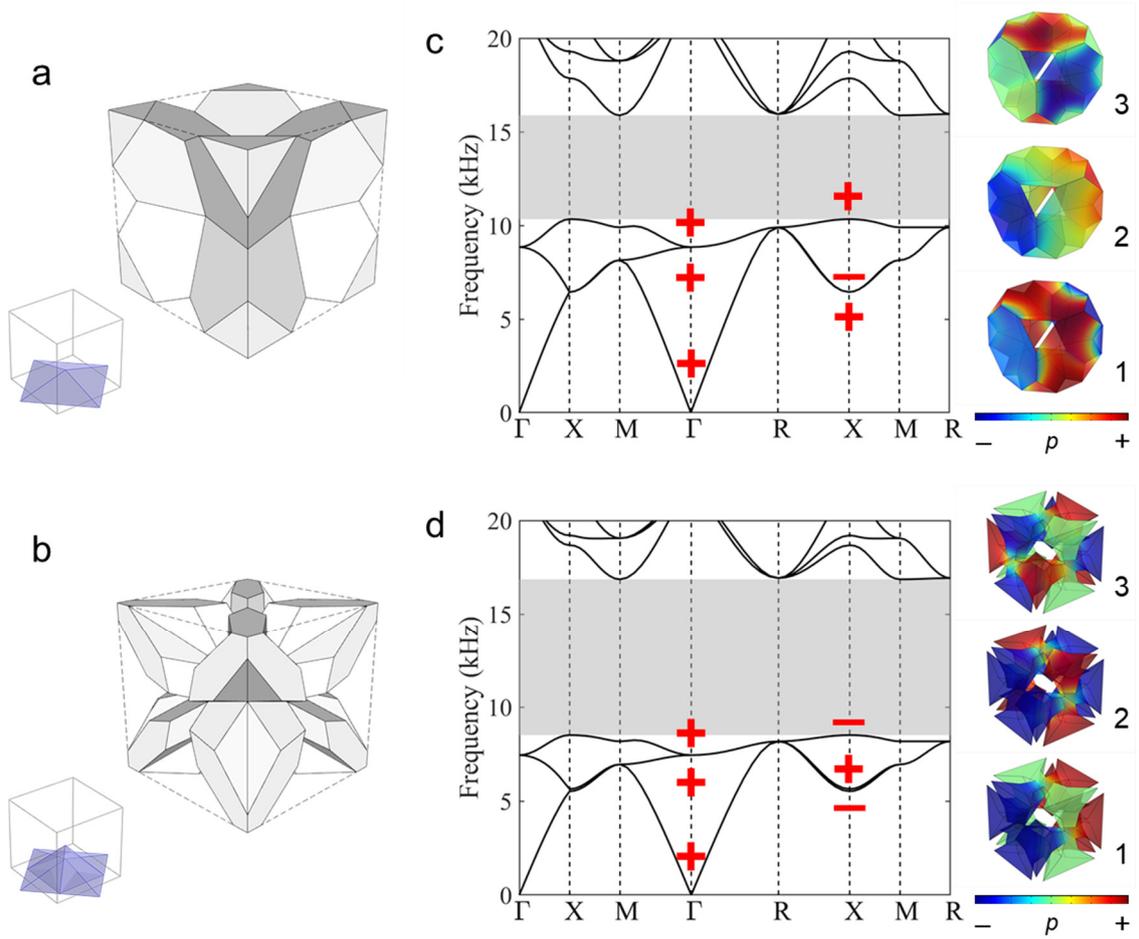

**Figure 1 Structure and bulk band diagrams of 3D sonic crystals.** (a), (b) Structure of the unit cells for non-trivial and trivial sonic crystals. The insets show the volumes removed from one of the six faces (denoted by the blue color). The pyramid in (a) has a height of $0.2a$ and base length of $0.7\sqrt{2}a$, while in (b), one more pyramid with height $0.4a$ and base length $0.7a$ are removed. (c), (d) Band diagrams of the non-trivial and trivial sonic crystals. Their band gaps are 10.34~15.88 kHz and 8.51~16.86 kHz, respectively. The parities of the eigenstates preserve at Γ but reverse at X ("+" denotes even parity while "−" denotes odd parity). The insets show the sound pressure fields of the first three eigenstates at X.

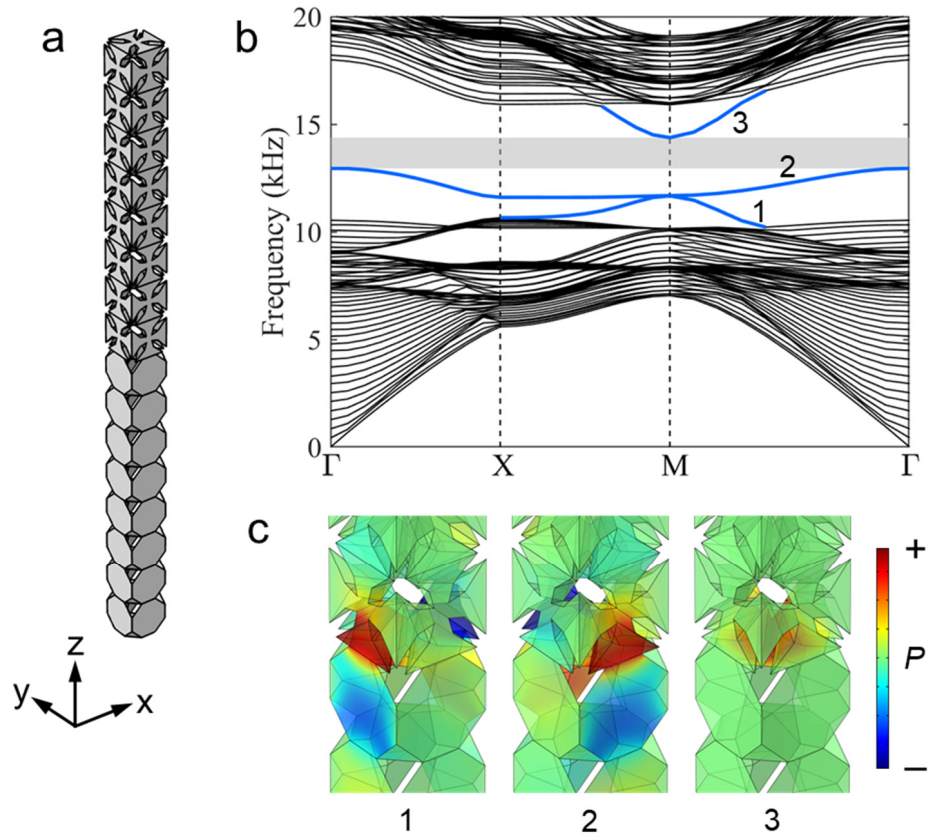

**Figure 2 Topological surface states.** (a) The ribbon-like supercell. Floquet-Bloch periodic boundary is applied in *x, y* direction, while the plane wave radiation boundary is applied on the top and bottom ends. (b) Projected band diagram of the surface supercell. Black lines denote bulk bands, while the blue lines denote surface bands. A surface band gap emerges between them. (c) Sound pressure profile for eigenstates at M of the surface bands. The left and middle ones are states at the two degenerated lower surface bands 1 and 2. The right one is the state at the highest surface band 3.

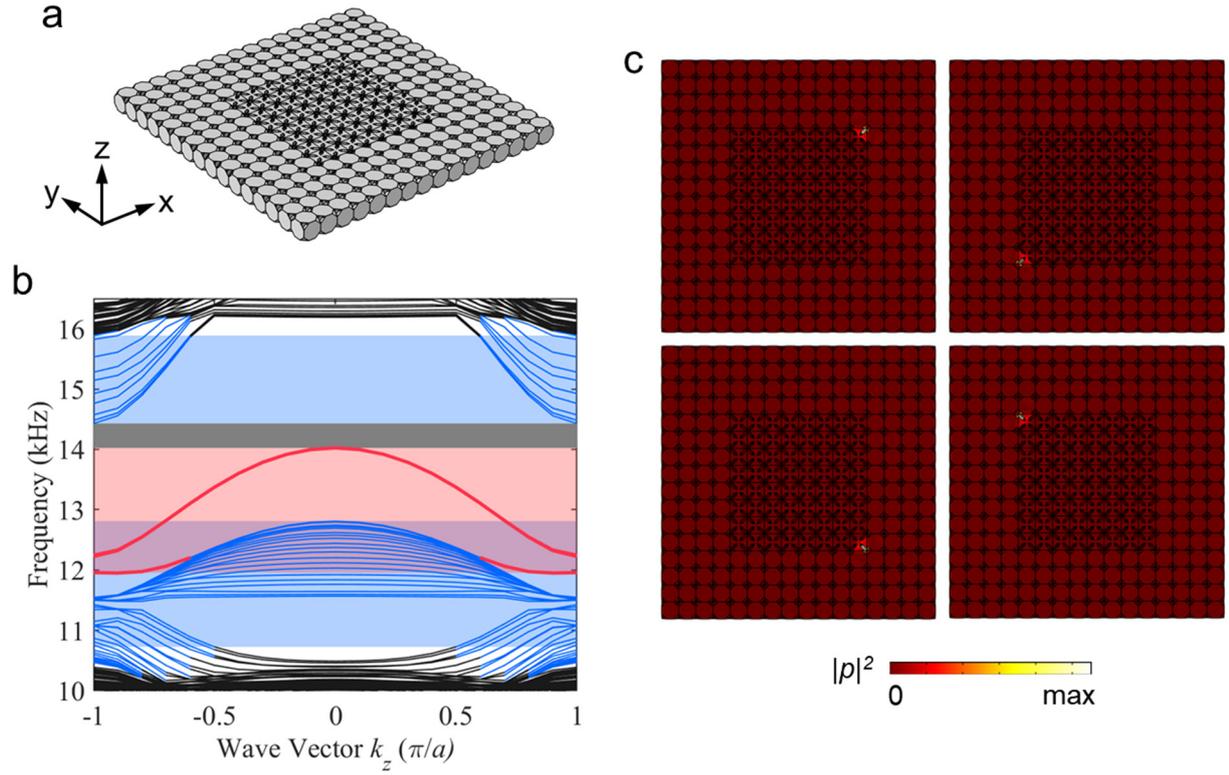

**Figure 3 Topological hinge states.** (a) The slab-like supercell. Floquet-Bloch periodic boundary is applied at the top and bottom faces, while the plane wave radiation boundary is applied at the other faces in *x* and *y* directions. (b) Projected band diagram of the supercell where hinge states are denoted by the red lines. The grey patch denotes the complete band gap for all propagation states, the red patch denotes the frequency range of hinge states, and the blue patch denotes surface states. Please note that there are four-fold hinge states around 12 kHz that merge into surface states when $|k_z|$ decreases. 11.96 to 12.81 kHz is the concurrent region of hinge and surface states. At below 10.72 kHz and above 15.89 kHz, the surface states gradually merge into bulk states when $|k_z|$ decreases. (c) Sound pressure field of 4 hinge states at 14.02 kHz ($k_z = 0$). The sound energy localizes at the four corners in the model.

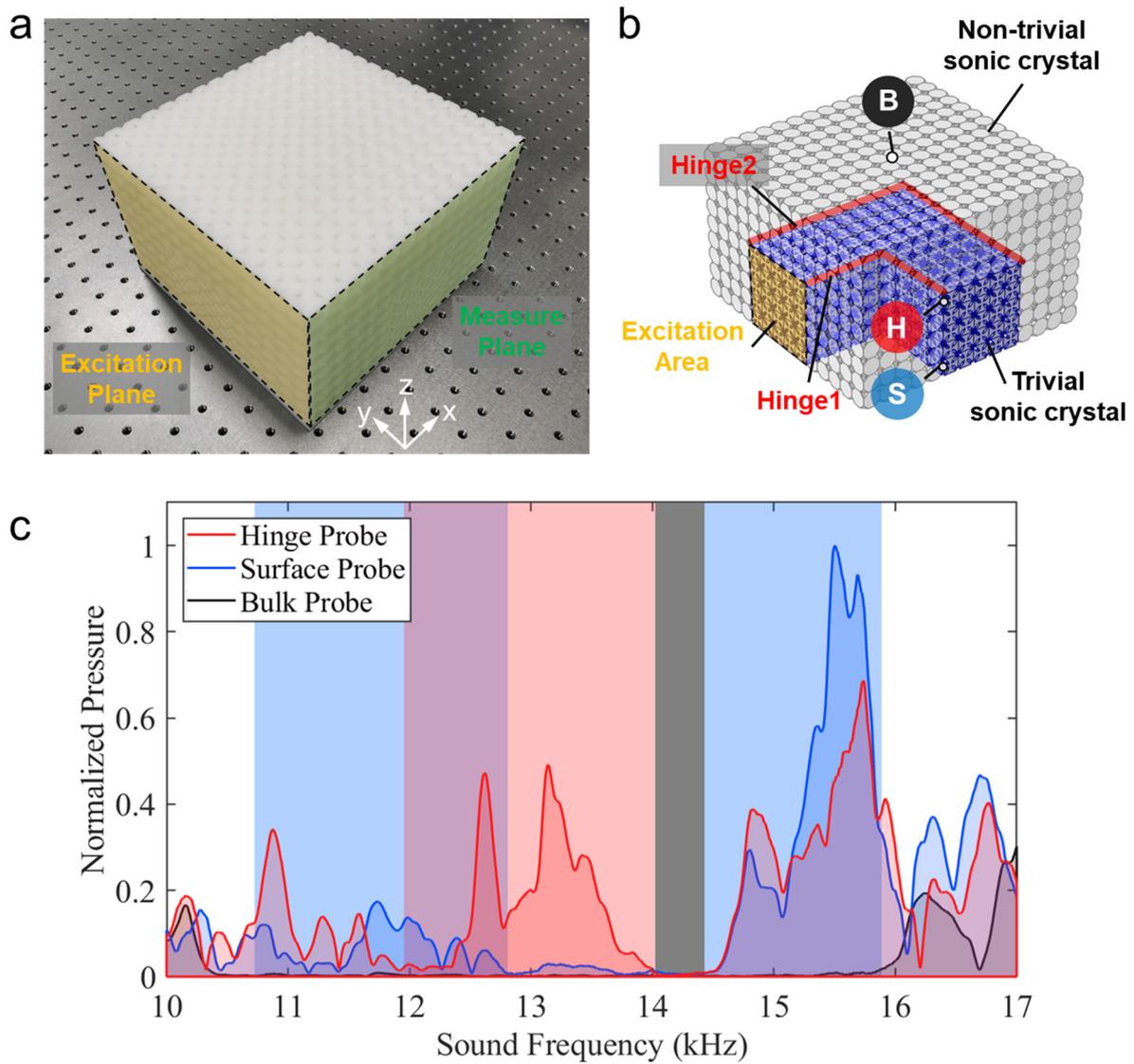

**Figure 4 Transmission spectra of multidimensional topological propagation states.** (a) Photograph of the fabricated sample. (b) Sketch of the experiment set up. The colored circles illustrate the position of microphones. "H", "S", "B" indicate hinge, surface, and bulk probe, respectively. The sound source is placed on the left side of the sample and exerted on the section of the trivial sonic crystal (yellow area). (c) Measured transmission spectra for the hinge (red), surface (blue), and bulk (black) probes. The colored patches indicate the numerically calculated frequency ranges of propagation states in Figure 3b.

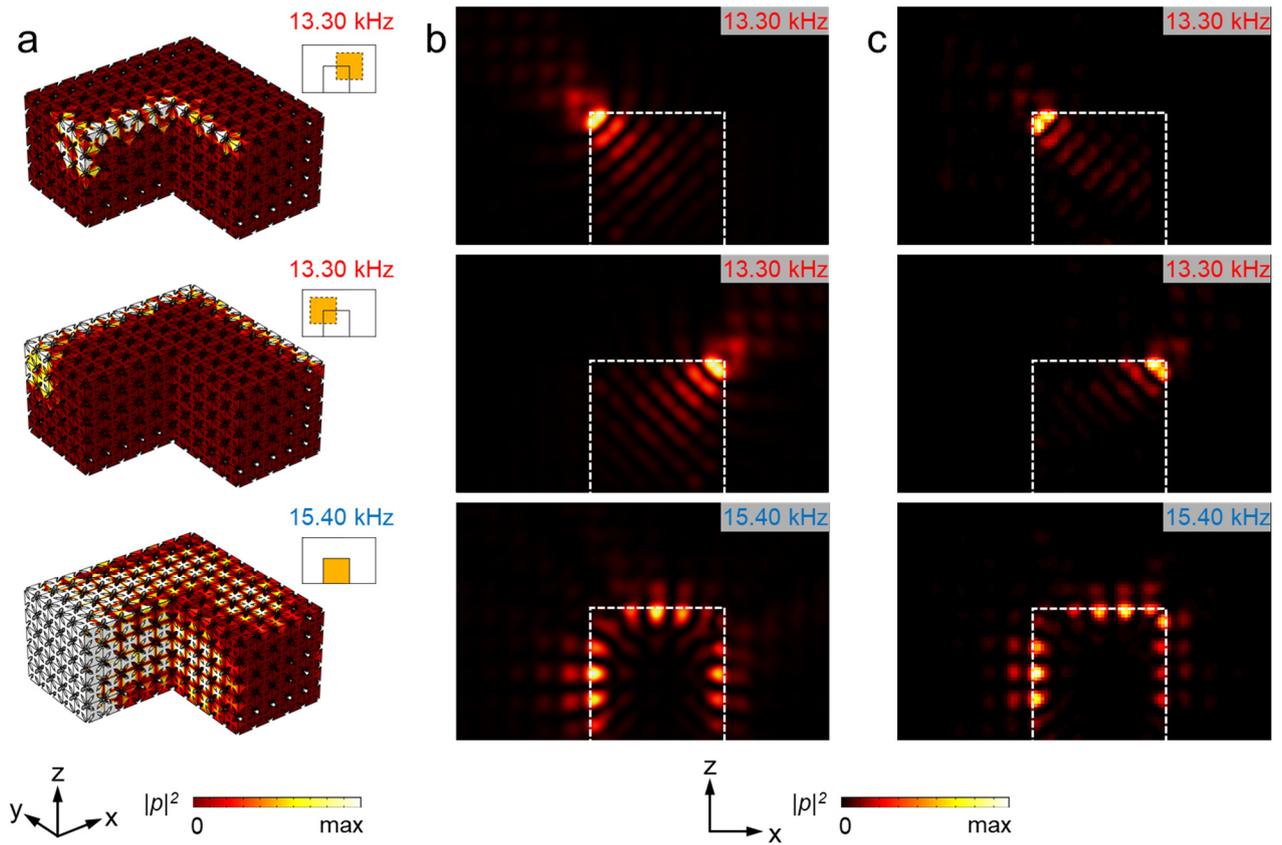

**Figure 5 Simulated and measured sound profile for surface state and hinge state.** The frequencies are 13.30 and 15.40 kHz, respectively. (a) Sound pressure profile inside the sample obtained from numerical simulation. The unit cells of non-trivial sonic crystal are hidden to present a clear view. The insets demonstrate the position of the sound source on the excitation plane in Figure 4a. (b) Simulated and (c) measured sound pressure profile at 1 mm away from the measure plane. The white dashed lines denotes the position of the interfaces between the trivial and non-trivial sonic crystals.

# Multidimensional sound propagation in 3D high-order topological sonic insulator

Fei Meng[1,2], Yafeng Chen[1,3], Weibai Li[1], Baohua Jia[1,*], Xiaodong Huang[1,*]

1 Centre of Translational Atomaterials (CTAM), Faculty of Science, Engineering and Technology, Swinburne University of Technology, Hawthorn, VIC 3122, Australia

2 State Key Laboratory of Geomechanics and Geotechnical Engineering, Institute of Rock and Soil Mechanics, Chinese Academy of Sciences, Wuhan, 430071, China

3 Key Laboratory of Advanced Technology for Vehicle Body Design & Manufacture, Hunan University, Changsha 410082, China

* Corresponding author: Baohua Jia, bjia@swin.edu.au; Xiaodong Huang, xhuang@swin.edu.au

# 1. 0D Corner State

The interface between the trivial and non-trivial sonic crystals in the main text is the platform for topological states in different dimensions, including 2D surface state, 1D corner state, and 0D corner state. To reveal the 0D corner state, we build a numerical model of a supercell, as shown in Figure S1a. The inner part of the supercell is the trivial sonic crystal with 8×8×8 unit cells. They are surrounded by 3 layers of non-trivial sonic crystal. Hence, 8 corners are formed in this supercell.

We calculated 600 eigenfrequencies around 13.50 kHz, as shown in Figure S1b. The blue dots are the topological surface states, black dots are bulk states, and red dots are hinge states. Range 14.02~14.43 kHz is a gap for all propagation modes according to Figure 3. However, 16 extra modes emerges in this region. The sound pressure field for a typical one of them is illustrated in Fig S1c. It can be seen that the sound energy localizes on the corners, which confirms that they are the topological corner states.

The frequencies of the 16 corner states do not exactly coincide but distribute in a small region from 14.219 to 14.253 kHz. The reason is that the number of unit cells for the inner part and out part is inadequate. For comparison, we build another supercell with 12×12×12 unit cells, and calculate 20 eigenfrequencies around 14.20 kHz. The result shows that the range of these corner states narrow down to 14.232 to 14.247 kHz. In the mode diagram, the sound energy also concentrates on the corners, as shown in Figure S1d. These corner states will gradually converge to the same frequency if we keep increasing the number of unit cells in the supercell. But currently, our computation resource (Quad Intel Xeon Gold 6128 CPU, 383GB RAM) can not afford such a simulation for this 3D topological insulator.

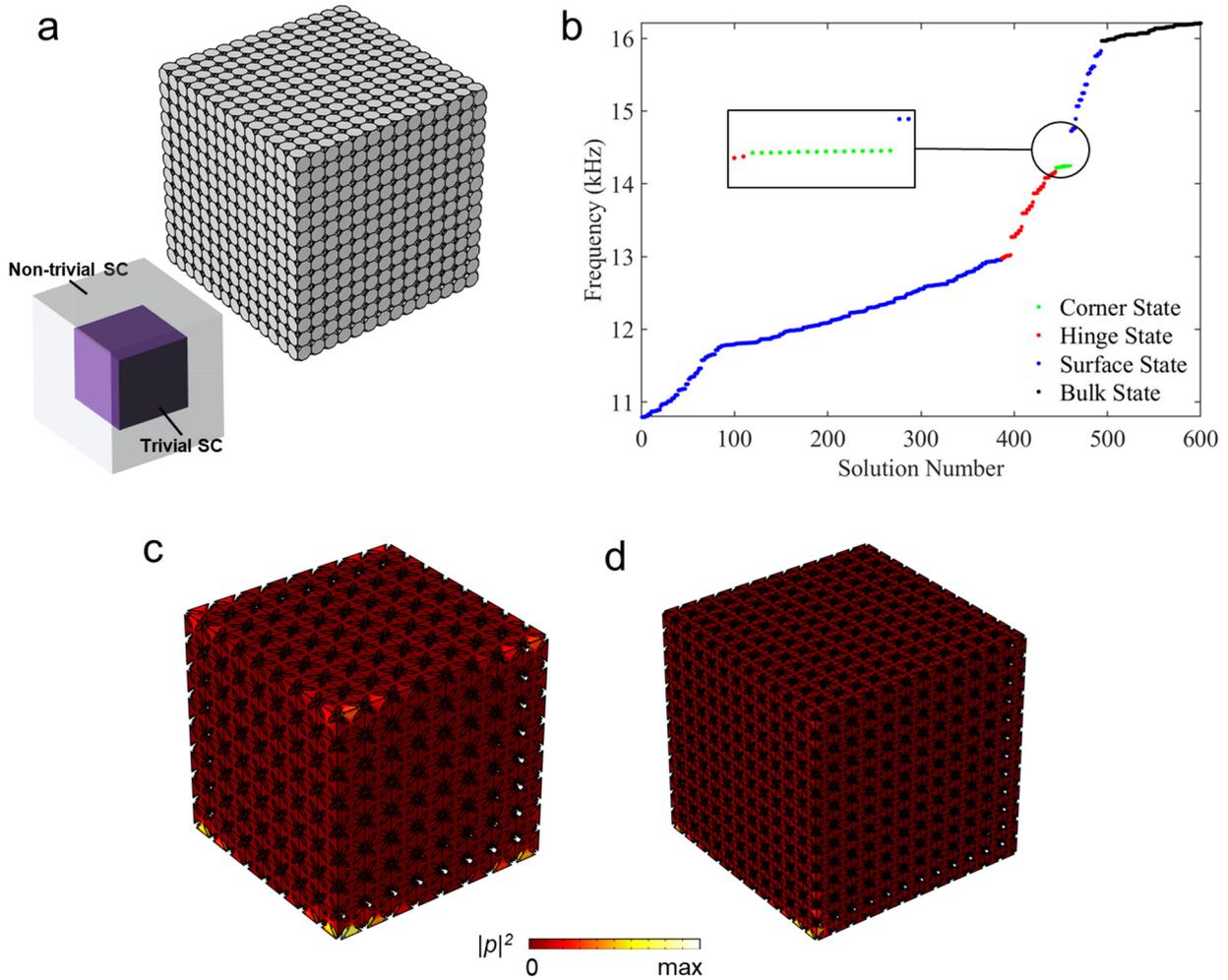

**Figure S1 Topological corner states.** (a) The cube-like supercell. The boundary condition is set as plane wave radiation boundaries in all three directions. (b) Numerical simulation result of the eigenfrequencies. (c), (d) Sound pressure field of corner states for the supercell with 8×8×8 (c) and 12×12×12 (d) unit cells of trivial sonic crystal inside. The non-trivial sonic crystals are hidden to present a clear view.

## 2. Transportation of sound in 3D

In the main text, we have numerically and experimentally demonstrate the transportation of sound in a sample through turning surfaces and hinges. In this section, we demonstrate a more complex numerical model. The trivial sonic crystal forms a square tunnel that turns twice in $y$ and $z$ direction. The side length of the tunnel is 5 unit cells. The trivial sonic crystal is surrounded by 5 layers of non-trivial sonic crystals, as shown in Figure S2a. The bottom, left, right and back side of the model are sealed and hence set as sound hard boundaries, while the other sides of the model are set as plane wave radiation boundaries. The sound source is exerted on the front section of the square tunnel.

At 13.30 kHz and 15.40 kHz, the sound pressure profile inside the model is revealed as in Figure S2b. It can be seen that the sound propagates through the hinge and the surfaces, respectively. The sound pressure fields at 1 mm away from the top surface are shown in Figure S2c. At 13.3 kHz, the sound pressure is highly localized at the exit of the hinge, while at 15.4 kHz, the acoustic energy concentrates on the interfaces. This simulation verifies the 1D hinge states and 2D corner states realized in this paper can be used to transport sound in 3D space, like in a 3D "strong" topological insulator.

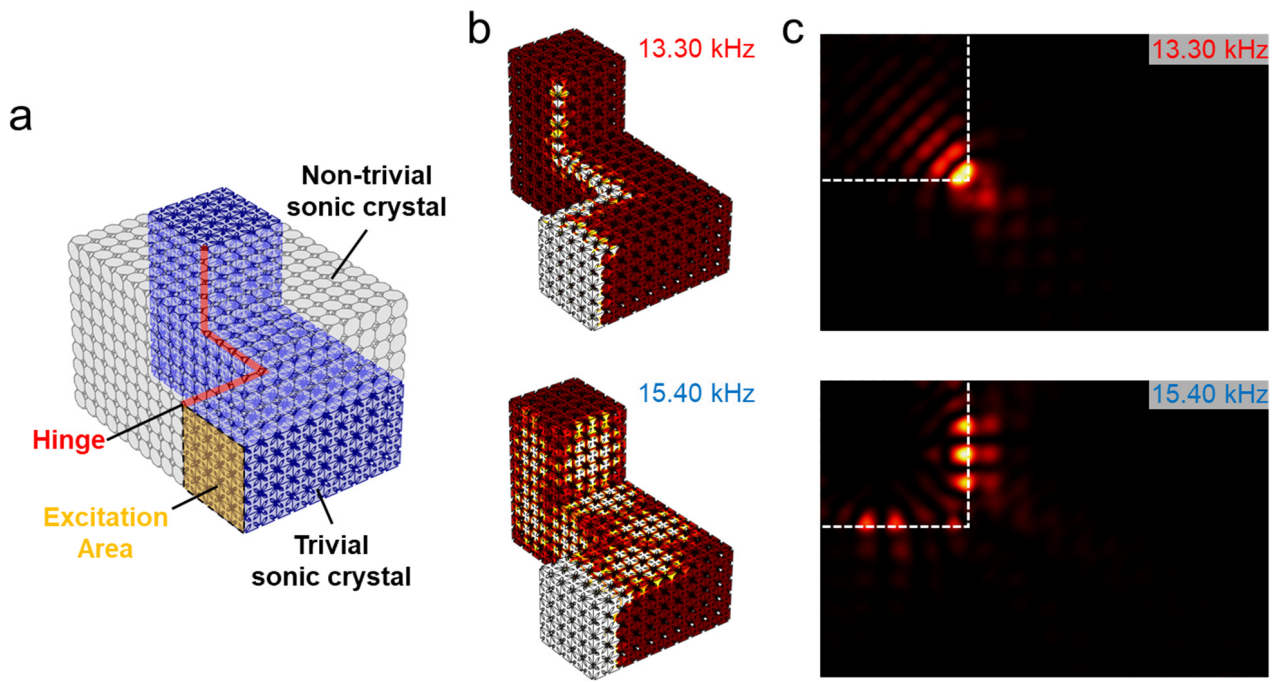

**Figure S2. A numerical model with curved hinge and surfaces.** The frequencies are 13.30 and 15.40 kHz, respectively. (a) Sketch of the numerical model. (b) Sound pressure profile inside the sample obtained from numerical simulation. The unit cells of non-trivial sonic crystal are hidden. (c)

Simulated sound pressure profile at 1 mm away from the top surface. The white dashed lines denotes the position of the interfaces between the trivial and non-trivial sonic crystals.